\documentclass[aps,prb,fleqn,12pt,showpacs,superscriptaddress]{revtex4}
\usepackage{epsfig,color}
\usepackage{graphicx}
\usepackage{pdfpages}
\usepackage{braket}
\usepackage{amssymb,amsmath}
\usepackage{amsfonts}
\usepackage{hyperref} 
\begin{document}
\title{Magnetic order and anisotropic interactions \\ induced by mixing between the $J=1/2$ and $3/2$ sectors \\ in spin-orbit coupled honeycomb-lattice compounds}
\author{Shubhajyoti Mohapatra}
\affiliation{Department of Physics, Indian Institute of Technology, Kanpur - 208016, India}
\author{Avinash Singh}
\email{avinas@iitk.ac.in}
\affiliation{Department of Physics, Indian Institute of Technology, Kanpur - 208016, India}
\date{\today} 
\begin{abstract}
Novel magnetic ordering on the honeycomb lattice due to emergent weak anisotropic interactions generated by the mixing between the $J=1/2$ sector and the magnetically inactive 3/2 sector is investigated in a three-orbital interacting electron model in the absence of Hund's coupling. Self-consistent determination of magnetic order yields anisotropic N\'{e}el and zigzag orders for different parameter regimes, highlighting the effect of the emergent single-ion anisotropy. Study of magnon excitations shows extremely small magnon energy scale compared to the hopping energy scale, and enhancement of anisotropy effects for smaller spin-orbit coupling. These results account for several features of the honeycomb lattice compounds such as $\rm Na_2 Ir O_3$ and $\rm Ru Cl_3$, where the leading order anisotropic interactions within the magnetically active $J=1/2$ sector are completely quenched due to the edge-sharing octahedra.
\end{abstract}
\maketitle

\newpage
\section{Introduction}

The $5d$ and $4d$ transition-metal honeycomb lattice compounds such as $\rm Na_2 Ir O_3$ and $\alpha$-$\rm Ru Cl_3$ have attracted renewed attention recently due to the observed novel magnetic ground state and fingerprints of proximate spin-liquid behavior inferred from the magnetic excitations.\cite{banerjee_SC_2017,revelli_arxiv_2019} These compounds are magnetic insulators with collinear zigzag antiferromagnetic (AFM) order below $7-15$ K, as confirmed by both resonant magnetic X-ray scattering and neutron scattering experiments.\cite{liu_PRB_2011,choi_PRL_2012,ye_PRB_2012,sears_PRB_2015,johnson_PRB_2015} Inelastic neutron scattering (INS) studies have identified very low-energy magnon modes below 10 meV in both compounds, with magnon gap of $\sim$ 2 meV in $\rm Ru Cl_3$.\cite{choi_PRL_2012,ran_PRL_2017} Resonant inelastic X-ray scattering (RIXS) in $\rm Na_2 Ir O_3$ also shows magnetic excitations extending up to 35 meV, with additional peaks in the range 0.4 to 0.8 eV associated with excitonic modes.\cite{gretarsson_PRB_2013} 
In both systems, the N\'{e}el temperature is low compared to the Curie-Weiss temperature ($\sim$ −100 K),\cite{choi_PRL_2012,sears_PRB_2015} supported by the recent finding of nearest neighbour spin correlation surviving well beyond the magnetic ordering temperature.\cite{banerjee_SC_2017,revelli_arxiv_2019} 

Treating the $\rm Ir^{4+}$/$\rm Ru^{4+}$ ions in $d^5$ configuration as spin-orbit coupled $J$=1/2 magnetic ions (neglecting the $J=3/2$ states), several theoretical spin models have been proposed to account for the magnetic order and low-lying excitations. These include the nearest neighbor Kitaev-Heisenberg (KH) model,\cite{chaloupka_PRL_2010} the KH-$J_2$-$J_3$ model,\cite{kimchi_PRB_2011,choi_PRL_2012,chkim_PRL_2012} and the KH model including symmetric off-diagonal (SOD) interactions.\cite{yamaji_PRL_2014,katukuri_NJP_2014,rau_PRL_2014,sizyuk_PRB_2014} Phase diagram obtained using these models show that the zigzag order is realized in a narrow parameter regime. Strong coupling expansion carried out in the $t_{\rm 2g}$ manifold yields inter-site anisotropic interactions in the $J=1/2$ spin model, which are explicitly proportional to the Hund's coupling term $J_{\rm H}$.\cite{rau_PRL_2014,hskim_PRB_2015,wang_PRB_2017} Thus, both the zigzag order and preferred spin orientation have been accounted for in terms of the inter-site anisotropic interactions between the $J$=1/2 ions only.

Density functional theory (DFT) investigations of electronic structures and corresponding tight-binding parameters for $\rm Na_2 Ir O_3$ and $\alpha$-$\rm Ru Cl_3$ have shown significant sensitivity to structural details such as monoclinic and trigonal distortions and presence of Na cation.\cite{foyevtsova_PRB_2013,hskim_PRB_2015,winter_PRB_2016,hskim_PRB_2016,wang_PRB_2017} 
Thus, in $\rm Na_2 Ir O_3$, the O-assisted nearest neighbor (NN) hopping term $t_2$ $(t_{pd\pi})$ is dominant over the hopping terms $t_1$ $(t_{dd\pi\delta})$ and $t_3$ $(t_{dd\sigma})$ arising from direct overlap of $d$ orbitals.\cite{foyevtsova_PRB_2013} Whereas, in $\alpha$-$\rm Ru Cl_3$, $t_3$ is nearly twice as large as $t_2$.\cite{hskim_PRB_2015} As for the preferred spin orientation in $\rm Na_2 Ir O_3$, both experimental as well as DFT studies are inconsistent in assigning ordered moment direction.\cite{singh_PRB_2010,liu_PRB_2011,choi_PRL_2012,ye_PRB_2012,chun_NAT_2015,hu_PRL_2015,gordon_JCP_2016}

Magnetic order and collective excitations in $\rm Na_2 Ir O_3$ have been investigated within a three-orbital interacting electron model in the Hartree-Fock (HF) and random phase (RPA) approximations.\cite{igarashi_JPCM_2016} With the staggered moment constrained along the crystal $a$ axis in the self-consistent analysis, zigzag order was obtained for weak direct $d$-$d$ hopping. Magnetic excitations in the zigzag state have also been investigated recently within the one-band Hubbard model with spin-dependent hopping terms.\cite{honecomb_JMMM_2019} However, fully unrestricted  self-consistent determination of magnetic order in the three-orbital model has not been carried out.

In the honeycomb lattice compounds such as $\rm Na_2 Ir O_3$ and $\rm Ru Cl_3$ with edge-sharing octahedra, there is a pair of orbital mixing hopping terms $t^{yz|xz}$ and $t^{xz|yz}$ for the $z$ bond (and similarly for the $x$ and $y$ bonds) corresponding to the two O (Cl)-assisted hopping pathways for each NN pair of Ir (Ru) ions. The two hopping terms in each pair are identical in the ideal cubic setting. Consequently, there are no spin-dependent hopping terms in the $J=1/2$ sector as the two hopping terms cancel each other (see Appendix A), and hence no leading-order anisotropic magnetic interactions in the $J=1/2$ sector. Furthermore, for $2t_1+t_3=0$, the usual (spin-independent) hopping terms also cancel, resulting in no isotropic Heisenberg interaction either in the $J=1/2$ sector.

However, the same orbital mixing hopping terms give rise to spin-dependent hopping terms between the $J=1/2$ and 3/2 sectors, which can effectively generate weak anisotropic magnetic interactions in the $J=1/2$ sector. Non-perturbative determination of these weak anisotropic interactions and the resulting novel magnetic orders have not been studied earlier. In this paper, we will therefore carry out self-consistent determination of magnetic order withn a three-orbital interacting electron model with spin-orbit coupling. Different magnetic orders such as cubic N\'{e}el, planar zigzag, and axial zigzag will be shown to be stabilized in different parameter regimes. 

The structure of this paper is as below. After introducing the three-orbital interacting electron model and transformation to the spin-orbit coupled $J$ basis states in Sec. II, the self-consistent determination of magnetic order is discussed in Sec. III. Magnon excitations in the different magnetic orders are investigated in Sec. IV, and characteristic features such as magnon gap and dispersion are related to the anisotropic interactions determined microscopically in Sec. V. Stability of the different magnetic orders is also discussed here in terms of minimal spin models. Some conclusions are presented in Sec. VI. 

\section{Three-orbital model and electronic band structure}

We consider a three-orbital interacting electron model in the $t_{2g}$ basis:
\begin{equation}
\mathcal{H} = \mathcal{H}_{\rm SOC} + \mathcal{H}_{\rm hop} + \mathcal{H}_{\rm int}
\label{three_orb_matrix}
\end{equation}
including the spin-orbit coupling (SOC), hopping, and Coulomb interaction terms, which will be discussed individually in the following subsections.

\subsection{SOC term and pseudo-orbital basis}
In the three orbital basis ($yz\sigma,xz\sigma,xy\bar{\sigma}$), the SOC term: 
\begin{equation}
\mathcal{H}_{\rm SOC} = \frac{\lambda}{2} \sum_{i,\sigma} \Psi_{i\sigma} ^{\dagger} \begin{pmatrix}
0 & i \sigma & -\sigma \\
- i \sigma & 0 & i \\
-\sigma & - i & 0 \\
\end{pmatrix} \Psi_{i\sigma}
\end{equation}
where $\lambda$ is the SOC strength and $\Psi_{i\sigma} ^\dagger$ = ($a_{i yz \sigma} ^{\dagger} \; a_{i xz \sigma} ^{\dagger} \; a_{i xy \bar{\sigma}} ^{\dagger}$) in terms of the creation operator $a_{i\mu \sigma} ^{\dagger}$ for site $i$, orbital $\mu=yz,xz,xy$, and spin $\sigma$=$\uparrow,\downarrow$. 

Our subsequent analysis will be carried out using the three spin-orbital-entangled Kramers pairs $|J,m_j \rangle$ which are the eigenstates of the SOC term. These pairs will be referred to as {\em pseudo orbitals} ($l=1,2,3$) with two {\em pseudo spins} ($\tau= \uparrow,\downarrow$). In terms of the $t_{2g}$ basis, the $|J,m_j \rangle$ and the corresponding $|l,\tau \rangle$ states have the form:
\begin{eqnarray}
\ket{l=1, \tau= \sigma} &=& \Ket{\frac{1}{2},\pm\frac{1}{2}} = \left [\Ket{yz,\bar{\sigma}} \pm i \Ket{xz,\bar{\sigma}} \pm \Ket{xy,\sigma}\right ] / \sqrt{3} \nonumber \\
\ket{l=2, \tau= \sigma} &=& \Ket{\frac{3}{2},\pm\frac{1}{2}} = \left [\Ket{yz,\bar{\sigma}} \pm i \Ket{xz,\bar{\sigma}} \mp 2 \Ket{xy,\sigma} \right ] / \sqrt{6} \nonumber \\
\ket{l=3, \tau= \bar{\sigma}} &=& \Ket{\frac{3}{2},\pm\frac{3}{2}} = \left [\Ket{yz,\sigma} \pm i \Ket{xz,\sigma}\right ] / \sqrt{2} 
\label{jmbasis}
\end{eqnarray}
where $\pm$ correspond to spins $\sigma = \uparrow/\downarrow$. Inverting the above transformation, we obtain the $t_{2g}$ basis states:
\begin{equation}
\begin{pmatrix}
a_{yz \sigma}^\dagger \\ a_{xz \sigma}^\dagger \\ a_{xy \overline{\sigma}}^\dagger \end{pmatrix}
= \begin{pmatrix}
\frac{1}{\sqrt{3}} & \frac{1}{\sqrt{6}} & \frac{1}{\sqrt{2}} \\
\frac{i\sigma}{\sqrt{3}} & \frac{i\sigma}{\sqrt{6}} & \frac{-i\sigma}{\sqrt{2}} \\
\frac{-\sigma}{\sqrt{3}} & \frac{\sqrt{2}\sigma}{\sqrt{3}} & 0
\end{pmatrix}  \begin{pmatrix} a_{1 \tau}^\dagger \\ a_{2 \tau}^\dagger \\ a_{3 \tau}^\dagger
\end{pmatrix}
\label{t2g_to_j}
\end{equation}
in terms of the pseudo-orbital basis states $\ket{l\tau}$, where $\sigma = \uparrow/\downarrow$ and $\tau = \overline{\sigma}$. 

\begin{figure}
\vspace*{-0mm}
\hspace*{-0mm}
\psfig{figure=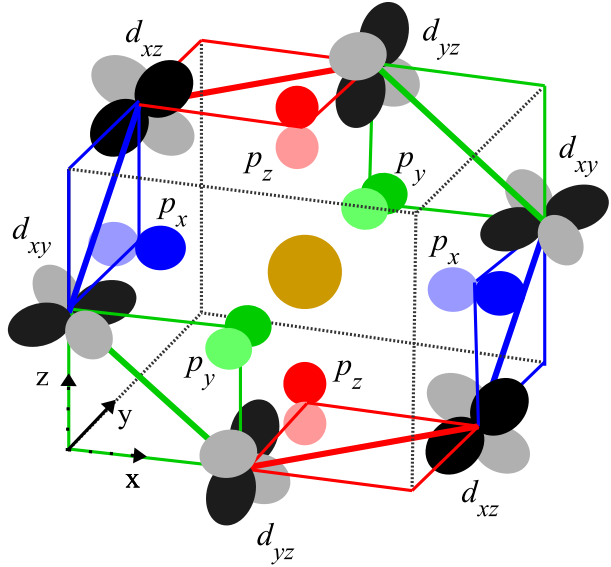,angle=0,width=80mm}
\vspace{-0mm}
\caption{Schematic representation of the honeycomb lattice structure of Ir/Ru ions in the cubic setting. For each of the three types of TM ion bonds, O(Cl)-assisted hopping involves mixing between two particular $t_{\rm 2g}$ orbitals: $d_{yz}-d_{xz}$ (red), $d_{xz}-d_{xy}$ (blue), and $d_{xy}-d_{yz}$ (green).} 
\label{embedded}
\end{figure} 

\subsection{Hopping terms}
\begin{figure}
\psfig{figure=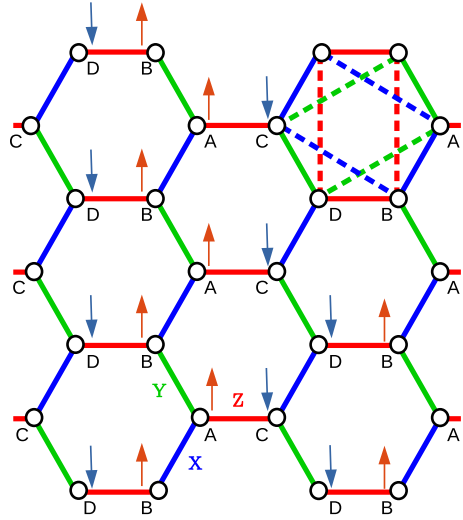,angle=0,width=60mm}
\caption{The $\rm{Z,X,Y}$ type bonds on the honeycomb lattice are shown in red, blue, green, with solid (dashed) lines for first (second) neighbors, along with the zigzag magnetic order and the four sublattices (A,B,C,D).} 
\label{lattice}
\end{figure}

In both $\rm Na_2 Ir O_3$ and $\rm Ru Cl_3$, the honeycomb lattice formed by $\rm Ir ^{4+}$/$\rm Ru ^{4+}$ ions is embedded in a cubic lattice, as shown in Fig. \ref{embedded}. The relevant processes that contribute to hopping mainly involve the two transition metal ions and their respective ligand octahedral cages. In the ideal cubic setting, inversion symmetry about the bond as well as time-reversal symmetry force the $[T_{ij}^{\gamma}]$ matrix to be both real and symmetric.\cite{rau_arxiv_2014} The hopping terms in the $t_{2g}$ basis are therefore given by:
\begin{eqnarray}
\mathcal{H}_{\rm hop}
& = & \sum_{\langle ij\rangle,\sigma} \Psi_{i\sigma}^{\dagger} 
[T_{ij}^{\gamma}] \Psi_{j\sigma}
\label{three_orb_two_sub}
\end{eqnarray} 
where $\gamma$=$X,Y,Z$ indicates the bond dependence, as shown in Fig. \ref{lattice}. By symmetry, the nearest neighbor hopping matrices for the $X,Y,Z$ bonds are given by:

\begin{eqnarray}
T_{ij} ^{Z} = \begin{pmatrix} t_1 & t_2 & 0 \\ t_2 & t_1 & 0 \\
0 & 0 & t_3 \end{pmatrix}, \;\;\; T_{ij} ^{X} =  
\begin{pmatrix}
t_3 & 0 & 0 \\ 0 & t_1 & t_2 \\ 0 & t_2 & t_1 \end{pmatrix}, \;\;\;
T_{ij} ^{Y} = \begin{pmatrix} t_1 & 0 & t_2 \\ 0 & t_3 & 0 \\
t_2 & 0 & t_1 \end{pmatrix}
\label{hopmat}
\end{eqnarray}
where $t_1$ and $t_3$ are the intra-orbital direct hoppings due to $\pi \delta$ and $\sigma$ overlaps, respectively, and $t_2$ is the inter-orbital indirect (O/Cl-assisted) hopping due to $\pi$ overlap. The trigonal and monoclinic distortion effect can be incorporated by including the orbital-mixing hopping term $t_4$ in place of zeroes in Eq. (\ref{hopmat}).\cite{rau_arxiv_2014} 

The second neighbor hopping is dominantly inter-orbital and again bond-dependent. This hopping occurs via two inequivalent channels.\cite{hskim_PRB_2015} In the first channel ($t_5$), hopping occurs between orbital lobes pointing towards honeycomb centre via intervening ligand $p$-orbitals or alkali metal $s$-orbital (if present at centre). In second channel ($t_5^\prime$), hopping occurs between orbital lobes pointing towards the intermediate TM site. The $t_5$ term is greater than $t_5^\prime$, and the two hopping terms are unequal due to absence of inversion symmetry about the bond.

Applying the transformation in Eq. (\ref{t2g_to_j}), the above hopping matrices are transformed to the pseudo-orbital basis of the $|l\tau\rangle$ states (Appendix A). The orbital mixing hopping term $t_2$ is only present in the off-diagonal blocks in the form of spin-dependent hopping terms $i\sigma_\mu t_\mu$ for the $\mu$=$X,Y,Z$ type bonds. As there are no spin-dependent hopping terms in the magnetically active $l$=1 ($J$=1/2) sector, only weak anisotropic magnetic interactions are induced due to mixing with magnetically inactive $l$=2,3 sectors. However, for the second neighbor hopping, the cancellation is avoided due to the two hopping channels being inequivalent, resulting in spin-dependent hopping and anisotropic interactions within the $J$=1/2 sector. 

It is well known that the honeycomb lattice yields a topological band insulator with four flat bands when only the O/Cl-assisted orbital-mixing hopping terms $t_2$ are included in the three-orbital model. 
The flat bands arise because of confined electron motion within a single honeycomb plaquette due to formation of quasi-molecular orbitals.\cite{mazin_PRL_2012,sohn_PRB_2013,li_PRB_2015} The SOC-induced mixing and splitting results in six bands with finite dispersion, arising from effective hopping between neighboring honeycomb plaquettes.  

\begin{figure}
\hspace*{0mm}
\psfig{figure=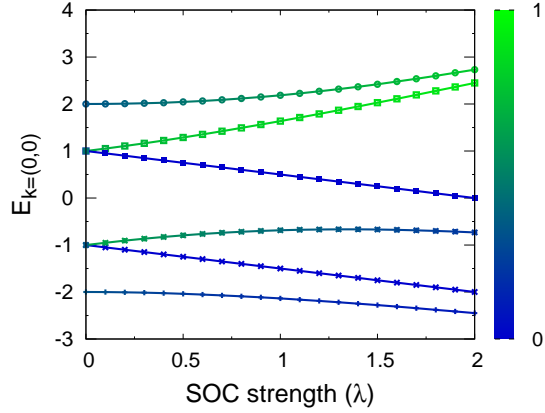,angle=-90,width=80mm}
\caption{Evolution of the electronic band energies at the $\Gamma$ point from dominantly quasi-molecular-orbital to $J$ character with increasing SOC strength $\lambda$ in the non-magnetic state. The colors green and blue indicate $J=1/2$ and 3/2 characters of the electronic states.} 
\label{evolution}
\end{figure}

Fig. \ref{evolution} shows the evolution of the electronic band energies at the $\Gamma$ point with increasing SOC value $\lambda$ in the non-magnetic state. The six band energies in the weak SOC limit progressively get regrouped with increasing SOC into the Kramer's doublet and quartet corresponding to $J$=1/2 and 3/2 states. For intermediate SOC strength, the electronic states will contain both quasi-molecular and $J$-state characters.\cite{bhkim_PRL_2016}

\subsection{Coulomb interaction and staggered field terms}
We consider the on-site Coulomb interaction terms in the three-orbital basis ($\mu,\nu = yz, xz, xy$) including the intra-orbital $(U)$ and inter-orbital $(U')$ density interaction terms, the Hund's coupling term $(J_{\rm H})$, and the pair hopping term $(J_{\rm H})$:
\begin{eqnarray}
\mathcal{H}_{\rm int} &=& U\sum_{i,\mu}{n_{i\mu\uparrow}n_{i\mu\downarrow}} + U^\prime \sum_{i,\mu < \nu,\sigma} {n_{i\mu\sigma}n_{i\nu\overline{\sigma}}} + \> (U^\prime - J_{\mathrm H}) \sum_{i,\mu < \nu,\sigma}{n_{i\mu\sigma}n_{i\nu\sigma}} \nonumber\\ & + & 
J_{\mathrm H} \sum_{i,\mu \ne \nu} \left ( {a_{i \mu \uparrow}^{\dagger}a_{i \nu\downarrow}^{\dagger}a_{i \mu \downarrow} a_{i \nu \uparrow}} + {a_{i \mu \uparrow}^{\dagger} a_{i \mu\downarrow}^{\dagger}a_{i \nu \downarrow} a_{i \nu \uparrow}} \right ) .
\end{eqnarray}
Here $a_{i\mu\sigma}^{\dagger}$ and $a_{i\mu \sigma}$ are the creation and annihilation operators for site $i$, orbital $\mu$, spin $\sigma=\uparrow,\downarrow$, and the density operator $n_{i\mu\sigma}$=$a_{i\mu\sigma}^\dagger a_{i\mu\sigma}$.

Using the change of basis (\ref{t2g_to_j}), the interaction Hamiltonian for site $i$ is transformed to the pseudo-spin-orbital basis:\cite{iridate_two}
\begin{eqnarray}
{\mathcal H}_{\rm int}(i) &=& \left( U - \frac{4}{3} J_{\rm H} \right) n_{1 \uparrow} n_{1 \downarrow} + \left( U - J_{\rm H} \right) \left[ n_{2 \uparrow}  n_{2 \downarrow} +  n_{3 \uparrow} n_{3 \downarrow} \right] \nonumber \\
&-& \frac{4}{3} J_{\rm H} {\bf S}_1 . {\bf S}_2 + 2 J_{\rm H} \left [ \mathcal{S}_{1}^z \mathcal{S}_{2}^z - \mathcal{S}_{1}^z \mathcal{S}_{3}^z \right] \nonumber \\  &+&
\left( U-\frac{13}{6} J_{\rm H} \right) \left[ n_{1} n_{2} + n_{1} n_{3} \right] + \left( U-\frac{7}{3} J_{\rm H} \right) n_{2} n_{3} 
\label{h_int}
\end{eqnarray} 
where we have used the spherical symmetry condition $U^\prime$=$U-2J_{\mathrm H}$. The pseudo-spin density operator ${\bf S}_{il} = \psi_{il} ^\dagger \frac{\mbox{\boldmath $\tau$}}{2} \psi_{il}$ and the charge density operator $n_{il} = \psi_{il} ^\dagger {\bf 1} \psi_{il} = n_{il \uparrow} + n_{il \downarrow}$ in terms of the Pauli matrices ${\mbox{\boldmath $\tau$}}$ and the local field operator $\psi_{il} ^\dagger = (a_{il\uparrow}^\dagger\;a_{il \downarrow}^\dagger)$. 

In order to focus on the emergent anisotropic interactions, we will neglect the Hund's coupling term $J_{\rm H}$ for simplicity. All three Hubbard-like terms ($n_{il\uparrow} n_{il\downarrow}$) then have the same interaction coefficient $U$. As the Hubbard-like terms ($\sim -{\bf S}_{il}.{\bf S}_{il}$) explicitly preserve spin rotation symmetry, the emergent anisotropic interactions arise only from the spin-dependent hopping terms which cannot be gauged away due to their bond-directional nature. 

We consider a four-sublattice basis (Fig. \ref{lattice}) in order to allow for N\'{e}el, zigzag, and stripy AFM orders, and discuss the staggered field terms arising from the HF approximation of the various interaction terms in Eq. (\ref{h_int}). For general AFM ordering with staggered field components {\boldmath $\Delta_{ls}$}=$(\Delta^x_{ls},\Delta^y_{ls},\Delta^z_{ls})$ for the three pseudo orbitals ($l$=1,2,3) and four sublattices ($s$=1-4), we obtain:  
\begin{equation}
\mathcal{H}^{\rm HF}_{\rm int} = \sum_{{\bf k}ls} \psi_{{\bf k}ls}^{\dagger} 
\begin{pmatrix} - \makebox{\boldmath $\tau . \Delta_{ls}$}
\end{pmatrix} \psi_{{\bf k}ls} 
= \sum_{{\bf k}ls} \psi_{{\bf k}ls}^{\dagger} 
\begin{pmatrix} -\Delta^z _{ls} & -\Delta^x _{ls} + i \Delta^y _{ls} \\
-\Delta^x _{ls} - i \Delta^y _{ls} & \Delta^z _{ls}  \\
\end{pmatrix} \psi_{{\bf k}ls}
\label{gen_ord_dirn} 
\end{equation}  
in ${\bf k}$ space, where $\psi_{{\bf k}ls}^\dagger=(a_{{\bf k}ls\uparrow}^\dagger\; \; a_{{\bf k}ls\downarrow}^\dagger)$. 

A composite 3-orbital$\otimes$4-sublattice$\otimes$2-spin basis will be employed to represent the HF Hamiltonian matrix with appropriate hopping terms in the ${\bf k}=(k_x,k_y)$ space. The staggered field components $\Delta_{ls}^\alpha$ (where $\alpha=x,y,z$) are self-consistently determined from:
\begin{equation}
2 \Delta^\alpha _{ls} = Um_{ls}^\alpha 
\label{selfcon}
\end{equation}
where the pseudo-spin magnetization components ${\bf m}_{ls}$=$m_{ls} ^x,m_{ls} ^y,m_{ls} ^z$ are evaluated using:
\begin{equation}
m_{ls}^\alpha = \sum_{{\bf k}\tau\tau'}^{E_{\bf k} < E_{\rm F}} \langle \varphi_{{\bf k}ls\tau} |
[\tau^\alpha]_{\tau\tau'} | \varphi_{{\bf k}ls\tau'} \rangle
\end{equation}
for the three pseudo-orbitals and four sublattices. Here $\varphi_{{\bf k}ls\tau}=\langle l\tau|\phi_{{\bf k}\mu s\sigma}\rangle$ are the eigenvectors of the HF Hamiltonian projected from the $t_{\rm 2g}$ basis to the pseudo-spin-orbital basis, and the summation is over all states below the Fermi energy. 

\section{Self-consistent determination of magnetic order}

\begin{table}
\caption{Self-consistently determined magnetic orders.} 
\centering 
\begin{tabular}{l l l l} \\  
\hline\hline 
Set \; & $t_1$, $t_2$, $t_3$, $t_4$, $t_5$ \hspace{15mm} & Magnetic order \;\; & $m_1^x$, $m_1^y$, $m_1^z$
\\ [0.5ex]
\hline 
A & -0.15, -1.0, 0.3, 0, 0 & cubic N\'{e}el & 0.48, 0.48, 0.51 \\[1ex]
B & -0.5, -0.5, 1.0, 0, 0 & planar zigzag & 0.63, -0.63, 0 \\[1ex]
C & -0.15, -1.0, 0.3, 0, 0.3 & axial zigzag & 0.17, 0.17, 0.86 \\[1ex]
D & -0.2, -1.0, 0.4, 0.15, 0 & planar zigzag & 0.60, -0.60, 0 \\[1ex]
\hline 
\end{tabular}
\label{table1}
\end{table}

For the self-consistent determination of the staggered field components, an iterative approach was employed starting with an initial choice for $(\Delta_{ls}^x,\Delta_{ls}^y,\Delta_{ls}^z)$ corresponding to zigzag order. In each iteration step, the local magnetization components were evaluated using the eigenvectors and eigenvalues of the HF Hamiltonian matrix, and the staggered field components were updated using Eq. (\ref{selfcon}). Typically, self consistency was achieved within few thousand iterations. 

We will consider four different sets of hopping parameters, as given in Table I. For the other parameters, we have taken $\lambda$=1.5, $U$=3.33 and $J_{\rm H}$=0, with the energy scale unit set by the largest hopping term. The usual hopping term $\sim (2t_1 + t_3)$ in the $l$=1 ($J$=1/2) sector (Appendix A) has been set to zero in all cases in order to suppress the isotropic Heisenberg interaction, and thus highlight the role of the emergent anisotropic interactions in determining the magnetic order. Also, in set C, we have taken $t_5'$=$t_5/2$. Both conditions are approximately consistent with DFT studies.   

The self-consistently determined magnetic orders, along with the magnetization components for the magnetically active sector $l$=1, are shown in the Table. Parameter sets A,C,D approximately correspond to $\rm Na_2 Ir O_3$, and with $t_2$=270 meV as obtained in DFT studies,\cite{foyevtsova_PRB_2013} realistic parameter values are obtained for the interaction term $U\approx 0.9$ eV and the SOC term $\lambda \approx 0.4$ eV. We find that the trigonal and monoclinic distortion effect, represented by the orbital-mixing hopping term $t_4$ (parameter set D), stabilizes the zigzag order instead of cubic N\'{e}el order as in set A, clearly showing the significant effect of structural distortion on the magnetic order. 

\begin{figure}
\psfig{figure=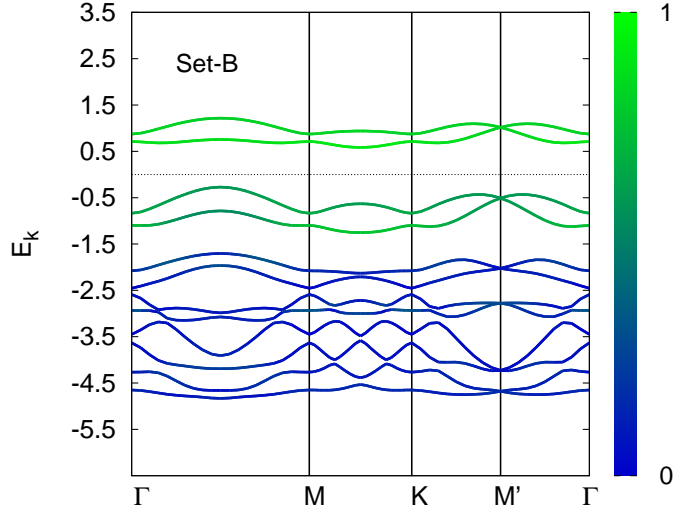,angle=-90,width=100mm}
\caption{Calculated electronic band structure for the zigzag order with parameter set B. The colors green and blue indicate $J=1/2$ and 3/2 characters of the electronic states.} 
\label{band}
\end{figure}

The electronic band structure is shown in Fig. \ref{band} for the planar zigzag order corresponding to parameter set B in Table I. All bands are weakly dispersive due to the significant quasi-molecular character associated with the confined electron motion within a single honeycomb plaquette. The states near the Fermi level (dotted line) have mostly $J$=1/2 character, and the band structure shows robust insulating gap in the magnetic state. The two conduction bands have stronger $J$=1/2 character than the two valence bands close to the Fermi energy. The calculated electronic band structure is in qualitative agreement with the DFT band structure.\cite{hskim_PRB_2015,koitzsch_PRL_2016}

With SOC reduced to half of the above value ($\lambda$=0.75) and interaction strength doubled ($U \approx 6$), set B hopping parameter values correspond more closely to the case of $\rm Ru Cl_3$ for which $t_3$=230 meV.\cite{hskim_PRB_2015} Although the anisotropy effects are significantly enhanced for smaller SOC due to stronger mixing between $J$=1/2 and 3/2 sectors, we find the same magnetic order as shown above. Effects of enhanced mixing on anisotropic magnetic interactions will be further discussed in the next section. 

\begin{table}
\caption{Magnetization values in the $J=3/2$ sector.} 
\centering 
\begin{tabular}{l l l l} \\  
\hline\hline 
Set \; & Magnetic order \; & $(m_2^x,m_2^y,m_2^z)$ \hspace{20mm} & $(m_3^x,m_3^y,m_3^z)$ \;\;
\\ [0.5ex]
\hline 
A & cubic N\'{e}el \;\; & (-0.004, -0.004, -0.02) & (-0.013, -0.013, 0.002) \\[1ex]
B & planar zigzag \;\; & (-0.007, 0.007, 0.0) & (0.034, -0.034, 0.0)  \\[1ex]
C & axial zigzag \;\; & (-0.008, -0.008, -0.04) & (0.014, 0.014, -0.04)  \\[1ex]
D & planar zigzag \;\; & (-0.02, 0.02, 0) & (0.03, -0.03, 0)  \\[1ex]
\hline 
\end{tabular}
\label{table2}
\end{table}

Table II shows the small magnetization values induced in the nominally filled $J$=3/2 sector ($l$=2,3) for the four parameter sets and magnetic orders. Since moments in the magnetically active $l$=1 sector (Table I) generate weak, anisotropic polarization in sectors $l$=2,3, these small moments feed back to the $l$=1 sector in the self-consistency process, and thus play a crucial role on the weak emergent anisotropic interactions. With increasing SOC and hence increasing spin-orbit gap, these magnetization values will progressively decrease and eventually vanish in the large SOC limit.

On including the Hund's coupling terms in Eq. (\ref{h_int}) in the HF approximation,\cite{iridate_two} self-consistent analysis for parameter set B yields axial zigzag order instead of planar zigzag order. Thus, instead of easy-plane anisotropy as obtained in $\rm Sr_2IrO_4$,\cite{iridate_two} easy-axis anisotropy is obtained here, the difference arising from the opposite sign of the significant moments in the $J$=3/2 sector. The Hund's coupling term $J_{\rm H}$ therefore effectively provides single-ion  anisotropy and mainly controls the zigzag ordering direction.  

\section{Magnon excitations}

The low-energy magnetic excitations were investigated using the time-ordered magnon propagator: 
\begin{equation}
\chi ({\bf q},\omega) = \int dt \sum_{i} e^{i\omega(t-t^\prime)}
e^{-i{\bf q}.({\bf r}_i - {\bf r}_j)}  
\langle \Psi_0 | T [ S_{i} ^{\alpha} (t) S_{j} ^{\beta} (t^\prime) ] | \Psi_0 \rangle
\label{chi}
\end{equation}
involving the $\alpha,\beta=x,y,z$ components of the $J=1/2$ spin operators $S_{i}^{\alpha}$ and $S_{j}^{\beta}$ at lattice sites $i$ and $j$. In the random phase approximation (RPA):
\begin{equation}
[\chi({\bf q},\omega)] = \frac{[\chi^0({\bf q},\omega)]}
{1 - 2U [\chi^0({\bf q},\omega)]}
\label{eq:spin_prop}  
\end{equation}
where the bare particle-hole propagator:
\begin{equation}
[\chi^0 ({\bf q},\omega)]_{s s'} ^{\alpha \beta} = \frac{1}{4} \sum_{{\bf k}} \left [ 
\frac{ 
\langle \varphi_{\bf k-q} | \tau^\alpha | \varphi_{\bf k} \rangle_s
\langle \varphi_{\bf k} | \tau^\beta | \varphi_{\bf k-q} \rangle_{s'} 
} 
{E^+_{\bf k-q} - E^-_{\bf k} + \omega - i \eta }
+ \frac{
\langle \varphi_{\bf k-q} | \tau^\alpha | \varphi_{\bf k} \rangle_s
\langle \varphi_{\bf k} | \tau^\beta | \varphi_{\bf k-q} \rangle_{s'} 
} 
{E^+_{\bf k} - E^-_{\bf k-q} - \omega - i \eta } \right ]
\label{chi0}
\end{equation}
was evaluated in the composite spin-sublattice basis (3 spin components$\otimes$4 sublattices) by integrating out the fermions in the self-consistently determined state. Here $E_{\bf k}$ are the eigenvalues of the HF Hamiltonian matrix in the three-orbital basis, the superscript $+(-)$ refers to particle (hole) energies above (below) the Fermi energy, and $\varphi_{{\bf k}\tau}$ are the projected amplitudes in the $J$=1/2 states. Magnon energies $\omega_{\bf q}$ are calculated from poles of Eq. (\ref{eq:spin_prop}). 

\begin{figure}
\hspace*{0mm}
\vspace{-20mm}
\psfig{figure=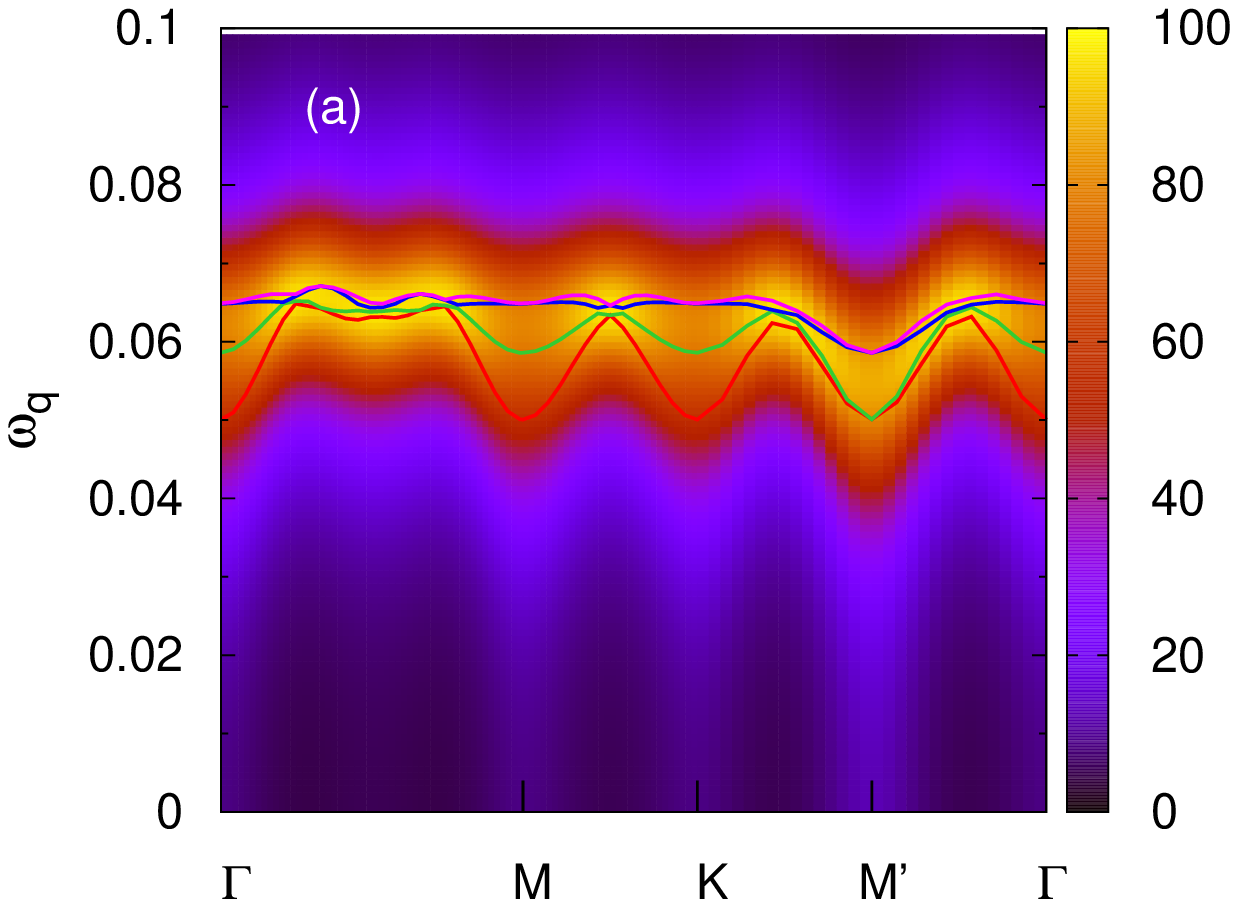,width=80mm}
\psfig{figure=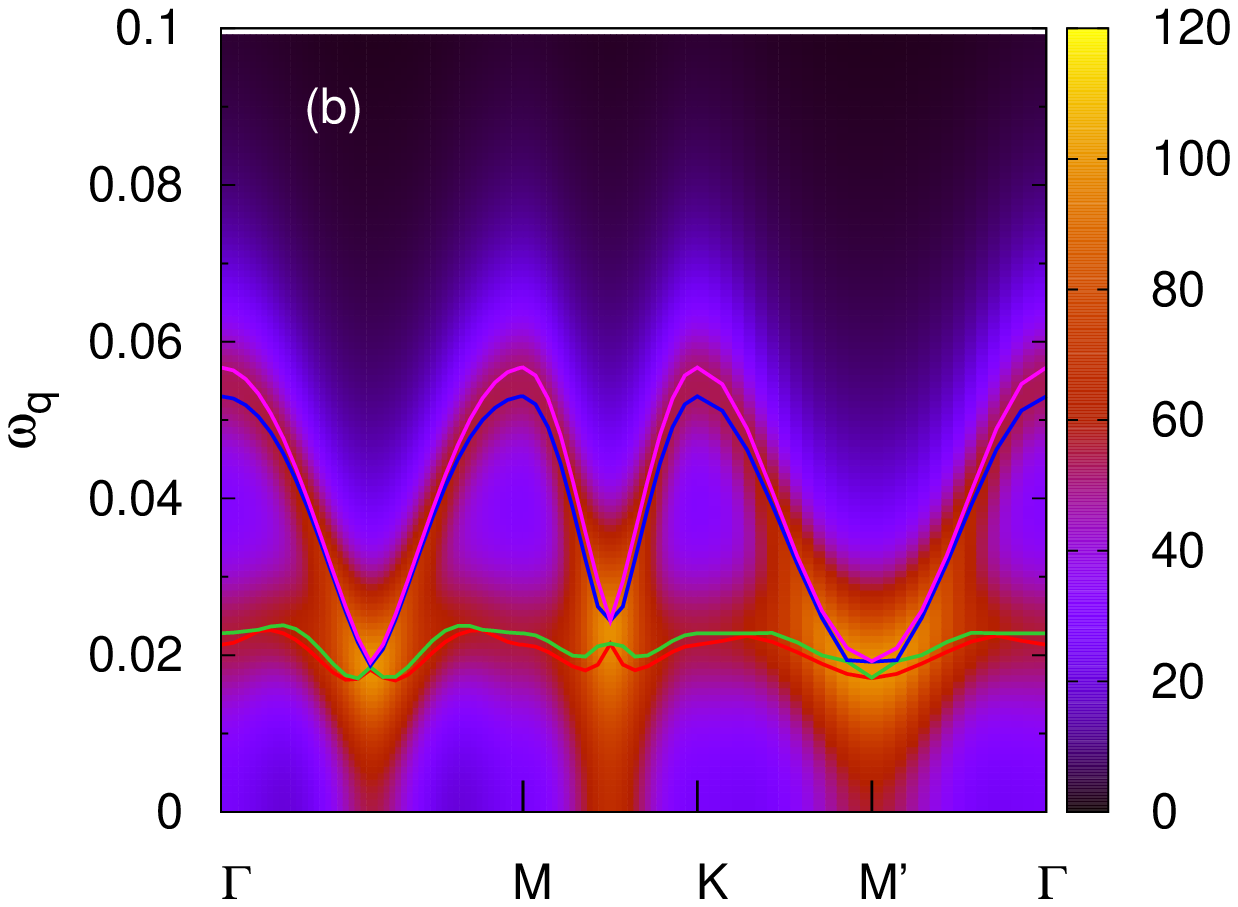,width=80mm} 
\vspace{-10mm}
\psfig{figure=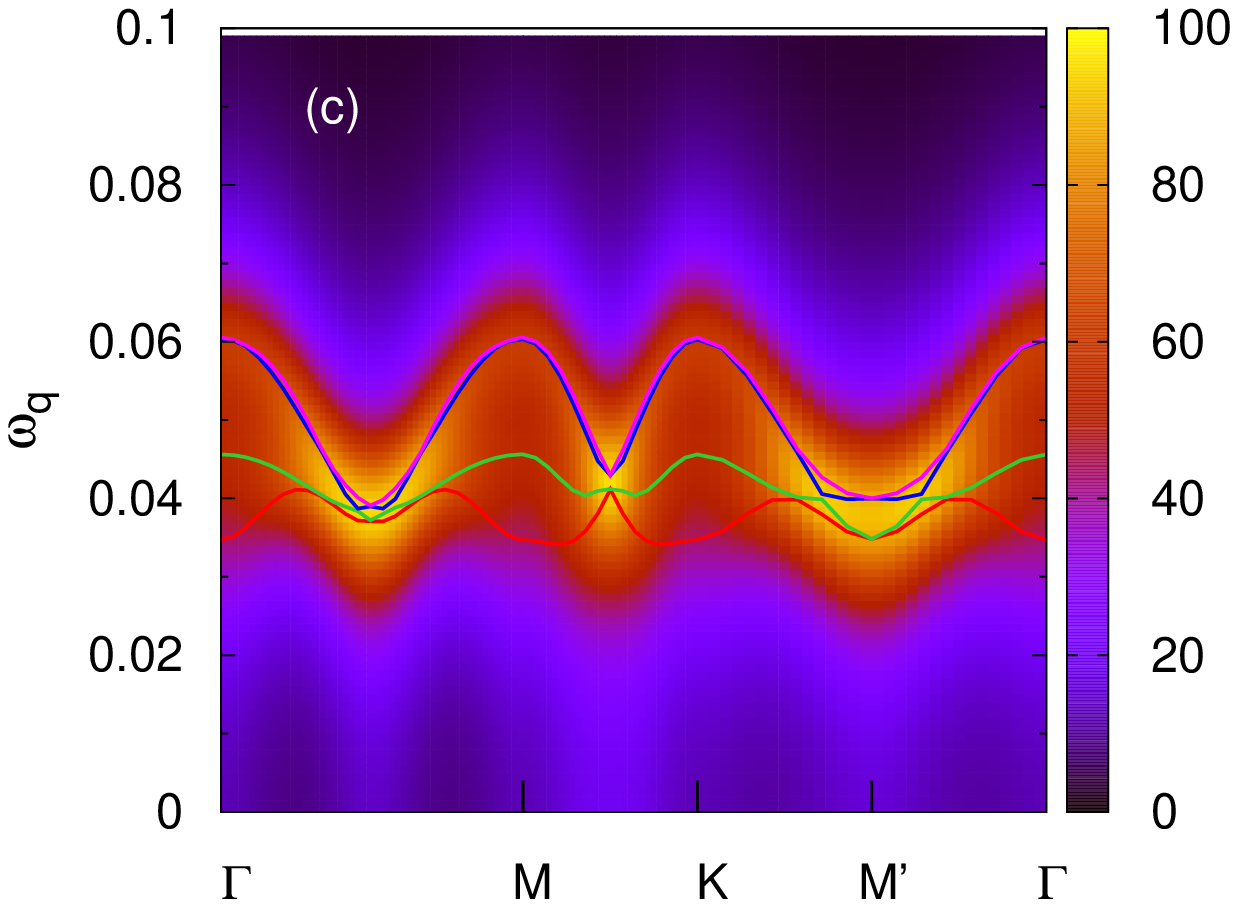,width=80mm}
\psfig{figure=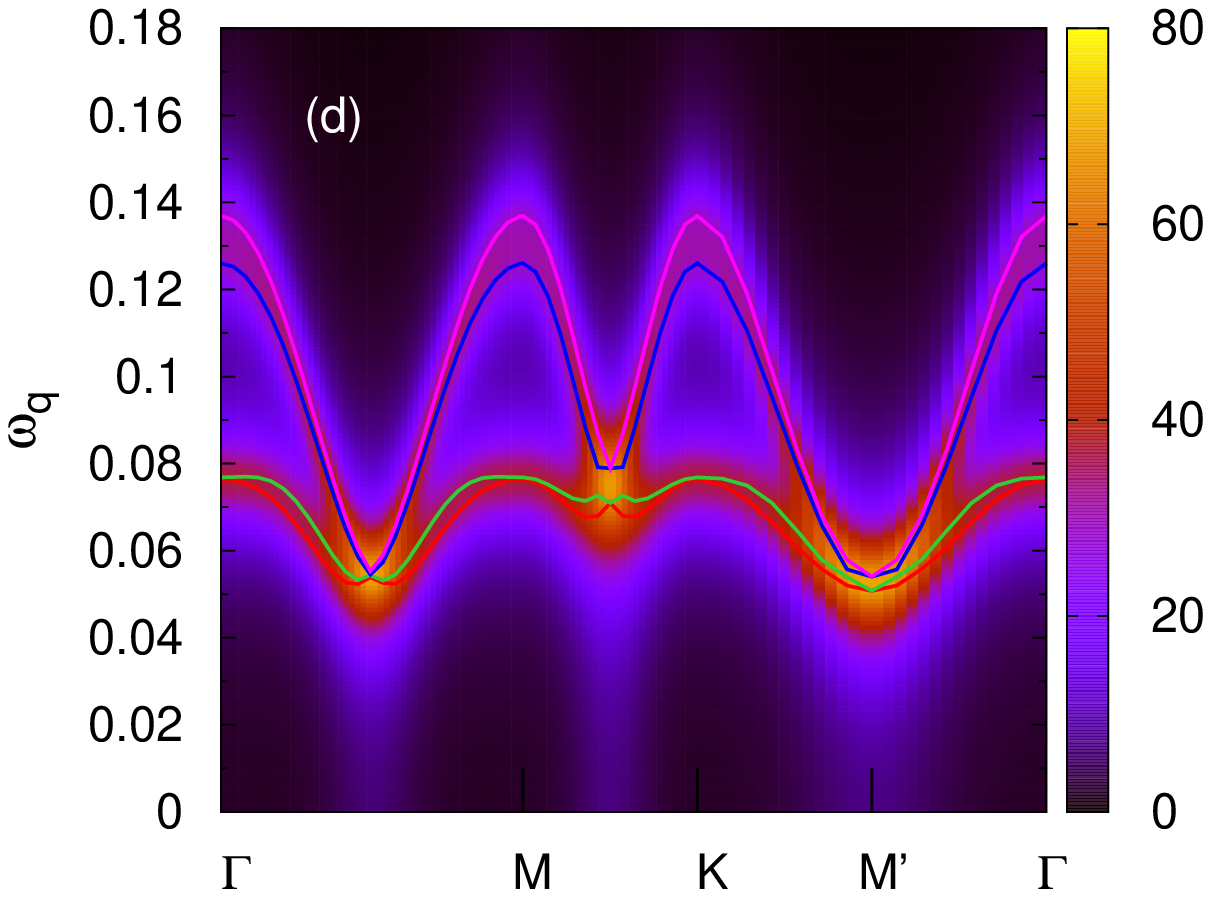,width=80mm} 
\caption{Calculated magnon spectral function and dispersion in the self-consistent state for different magnetic orders: (a) cubic N\'{e}el (parameter set A), (b,c) planar zigzag (sets B,D), and (d) planar zigzag (set B) with smaller SOC value $\lambda=0.75$.} 
\label{wq}
\end{figure}

Magnon dispersion and spectral function are shown in Fig. \ref{wq} for the self-consistently determined magnetic orders. Positive magnon energies over the entire Brillouin zone confirms that the magnetic orders obtained are stable and that the self-consistency process indeed yields the ground state in each case. The extremely small magnon energy scale as compared to the hopping energy scale is due to the weak emergent interactions generated only by the mixing with the magnetically inactive $J$=3/2 sector. 

Fig. \ref{wq}(a) for cubic N\'{e}el order clearly shows weak magnon dispersion around a large anisotropy gap, signifying that the intra-site interactions are dominant. On the other hand,   significantly broader dispersion for planar zigzag order (b) implies stronger inter-site interactions. The nearly flat, low-energy branches (b) indicate presence of intra-site interactions in the planar zigzag state as well. For realistic hopping energy scale $\approx$270 meV, the calculated magnon dispersion and gap are comparable to the experimental measurements in Ir/Ru based compounds.\cite{choi_PRL_2012,gretarsson_PRB_2013,ran_PRL_2017} Anisotropy effects are enhanced for smaller SOC (d), highlighting the non-perturbative role of mixing between the $J$=1/2 and 3/2 sectors. Also, the iteration process converges significantly faster for $t_2$=$-$0.7 in parameter set B (Table I), indicating more robust planar zigzag order, as also confirmed from enhanced magnon energies.  

\section{Emergent anisotropic magnetic interactions}

In the previous section, three type of magnetic orders were obtained which are stabilized by the weak anisotropic interactions generated effectively due to the spin-dependent hopping terms connecting sectors $l$=1,2,3. Due to this mixing, spin polarization (in some direction) in the magnetically active $l$=1 sector induces weak, anisotropic polarization (in other direction) in all sectors $l$=1,2,3, which evolve in the iteration process. Thus, for the parameter set A in Table I, starting with an initial axial zigzag order along the $z$ direction, the iteration process finally yields a self-consistent cubic N\'{e}el state with moments having $x,y,z$ components. Explicit evaluation of these anisotropic magnetic interactions is discussed below. 

In interacting electron models with purely local interactions, magnetic interactions are generated by the exchange of the particle-hole pair,\cite{ghosh_JPCM_2016} and are implicitly incorporated in the magnon propagator discussed above. This allows the anisotropic magnetic interactions to be determined microscopically from the ground state electronic band structure. If the generalized magnetic interaction is represented as:
\begin{equation}
H_{\rm spin} = \sum_{ij} \sum_{\alpha,\beta=x,y,z} S_{i \alpha} {\cal J}_{ij} ^{\alpha\beta} S_{j \beta}  
\end{equation}
where $S_{i\alpha}$ and $S_{j\beta}$ are the pseudo-spin operators (including both $J=1/2$ and 3/2 sectors), then the interaction terms can be evaluated from:
\begin{equation}
{\cal J}_{ij} ^{\alpha\beta} = -2U^2 [\chi^0]_{ij} ^{\alpha\beta} =-2U^2 \sum_{\bf q} [\chi^0({\bf q})]^{\alpha\beta} e^{i{\bf q}.({\bf r}_i - {\bf r}_j)} 
\label{couplings}
\end{equation}
in terms of the bare particle-hole propagator given in Eq. (\ref{chi0}) evaluated for $\omega=0$. The  above approach is well known to interpolate properly to the strong coupling limit, and is therefore particularly well suited for the intermediate coupling regime ($U \sim 1$ eV) relevant for the $5d^5$ and $4d^5$ honeycomb lattice compounds. Furthermore, this approach correctly yields the Kitaev interactions for the one-band Hubbard model on the honeycomb lattice with only spin-dependent hopping terms. 

Minimal spin models for each of the three magnetic orders are discussed below, keeping only the dominant anisotropic interactions. For the cubic N\'{e}el case, we found all three intra-site off-diagonal (OD) terms to be nearly equal, and the inter-site interactions to be negligible. The minimal spin model therefore includes only the intra-site interactions $-{\cal D} \left [S_{ix} S_{iy} + S_{iy} S_{iz} + S_{iz} S_{ix}\right]$, which are in conformity with the minimal requirement to stabilize the local (1,1,1) order. The three intra-site OD terms are contained within the generalized single-ion-anisotropy ${\bf S}_i . {\bf {\cal D}} . {\bf S}_i$ involving the interaction tensor ${\bf {\cal D}}$.\cite{neese_IC_1998,maurice_JCPC_2009}

For the planar zigzag case, only one intra-site OD interaction $+{\cal D}S_{ix}S_{iy}$ was obtained, which stabilizes the local $(1,-1,0)$ magnetic order. The inter-site interactions were found to be negligible for NN sites. However, for NNN sites, the interactions are finite and exhibit a combination of Kitaev, SOD, and DM terms, as seen from the ${\cal J}_{ij}^{\alpha\beta}$ matrix given in Appendix B. Retaining the dominant terms involving $x,y$ components only (as there is no $z$ moment), the interactions have the approximate form: $-C\left[S_{ix}S_{jx}+S_{iy}S_{jy}+S_{iy}S_{jx}/2 \right ]$ for the AA (Z) bond, $-\frac{C}{2} \left [ S_{ix} S_{jy} - S_{ix} S_{jx}/2 \right]$ for the AD (X) bond, and $-\frac{C}{2} \left [ (S_{iy} S_{jx} - S_{iy} S_{jy}/2 \right ]$ for the AD (Y) bond, where $C$ is a positive energy constant. 

For the axial zigzag case, intra-site OD interactions $-{\cal D}_\perp (S_{ix}+S_{iy})S_{iz}$ and $-{\cal D}_z (S_{iz})^2$ were obtained, which stabilize the local $(m_x,m_y,m_z)$ order with $m_x,m_y \ll m_z$. The inter-site interactions were again negligible for NN sites. The NNN interactions are dominant only for the AA (Z) bond, and have the approximate form: $-C'\left [S_{iz} S_{jz}/4 + (S_{ix} S_{jx} + S_{iy} S_{jy}) \right]$ corresponding to a combination of Kitaev and Heisenberg terms. As spins are parallel for the AA (Z) bond, all three terms yield negative energy and thus stabilize the magnetic order. Similarly, for the planar zigzag order, all terms (except $S_{iy}S_{jx}$) for all three bonds stabilize the magnetic order. 

\begin{figure}
\vspace*{-15mm}
\psfig{figure=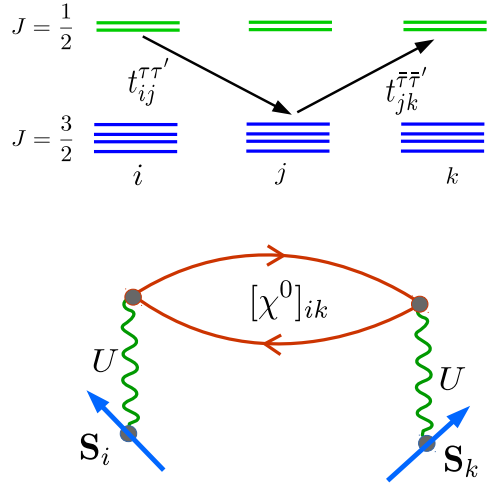,width=80mm} \vspace{-20mm}
\caption{Schematic diagram showing the effectively second-order hopping process involving the
NN spin-dependent hopping terms between the $J$=1/2 and 3/2 sectors (Appendix A), resulting in the anisotropic magnetic interaction generated by the exchange of the particle-hole pair.} 
\label{ph_pair}
\end{figure}

As discussed above, only NNN interactions between the $J$=1/2 moments were found to be induced, which is expected from an effectively second-order hopping process involving the NN spin-dependent hopping terms between the $J$=1/2 and 3/2 sectors (Fig. 6). However, NN interactions are also induced, but only between the $J$=1/2 and 3/2 moments, and are given in Appendix B for the planar zigzag order. Although their magnetic effects are strongly suppressed due to the extremely small $J$=3/2 moments, these NN interactions moderately frustrate the magnetic orders as seen from energetics. Extending the magnon calculation to include both $J$ sectors shows softening of magnon energies, which also supports this frustration effect. We note here that the $J$=3/2 moments are not proportional to the $J$=1/2 moments. Therefore, effective spin models involving only $J$=1/2 spins and NN anisotropic interactions (as considered in earlier works) will be insufficient to describe the competing interaction effects as found above. 

It should be emphasized that the anisotropic magnetic interactions are generated non-perturbatively in our three-orbital model even without the Hund's coupling term $J_{\rm H}$. In contrast, only inter-site anisotropic interactions were obtained perturbatively in earlier works using strong-coupling expansion,\cite{rau_PRL_2014} which are explicitly proportional to $J_{\rm H}$, and would therefore vanish for $J_{\rm H}=0$. 

\section{Conclusions}

Novel magnetic orderings on the honeycomb lattice were found to be stabilized in our non-perturbative investigation by the emergent weak anisotropic interactions induced only by the mixing with the magnetically inactive $J$=3/2 sector. Self-consistent determination of magnetic order yielded cubic N\'{e}el, planar zigzag, and axial zigzag states in different parameter regimes. The ordering directions were found to be locked with respect to the crystal axes due to the emergent single-ion anisotropy, which also accounted for the magnon gap in all cases. The weak anisotropic interactions were reflected in the extremely low magnon energies as compared to the hopping energy scale, and anisotropy effects were found to be enhanced for smaller SOC, as indicated by increased magnon energies. Structural distortion was found to significantly stabilize the zigzag order.  

Our non-perturbative approach provides insight into the mixing-induced anisotropic interactions not considered in earlier investigations. Only NNN interactions in the $J$=1/2 sector were found to be induced by the NN spin-dependent hopping terms between the $J$=1/2 and 3/2 sectors due to an effectively second-order hopping process. Having a combination of Kitaev, SOD, and DM terms, the anisotropic interactions were shown to stabilize the respective magnetic orders. NN anisotropic interactions induced between $J$=1/2 and 3/2 moments only were found to frustrate the magnetic orders, and therefore contribute to the proximate spin liquid regime. These effects are expected to be stronger for compounds with smaller SOC, such as $\rm RuCl_3$. 
  
\appendix 
\section{Transformed hopping matrices}
Applying the transformation (4), the hopping matrices (\ref{hopmat}) are transformed from the three-orbital ($t_{\rm 2g}$) basis to the pseudo-orbital basis $\{\ket {1\tau},\ket {2\tau},\ket {3\tau}\}$, where $\tau = \uparrow,\downarrow$:

\begin{eqnarray}
\tilde{T}_{ij} ^{Z}
&=& \left(
\begin{array}{cc|cc|cc}
\frac{2t_1 + t_3}{3} & 0 & \frac{\sqrt{2}(t_1 - t_3)}{3} & 0 & -i\frac{\sqrt{2}t_2}{\sqrt{3}} & 0 \\
0 & \frac{2t_1 + t_3}{3} & 0 & \frac{\sqrt{2}(t_1 - t_3)}{3} & 0 & i\frac{\sqrt{2}t_2}{\sqrt{3}}  \\
\hline
\frac{\sqrt{2}(t_1 - t_3)}{3} & 0 & \frac{t_1 + 2t_3}{3} & 0 & -i\frac{t_2}{\sqrt{3}} & 0 \\
0 & \frac{\sqrt{2}(t_1 - t_3)}{3} & 0 & \frac{t_1 + 2t_3}{3} & 0 & i\frac{t_2}{\sqrt{3}} \\
 \hline
i\frac{\sqrt{2}t_2}{\sqrt{3}} & 0 & i\frac{t_2}{\sqrt{3}} & 0 & t_1 & 0 \\
0 & -i\frac{\sqrt{2}t_2}{\sqrt{3}} & 0 & -i\frac{t_2}{\sqrt{3}} & 0 & t_1
 \end{array} \right), \nonumber \\
\ \\
\tilde {T}_{ij} ^{X}
 &=& \left(
\begin{array}{cc|cc|cc}
\frac{2 t_1+t_3}{3} & 0 & \frac{t_3-t_1}{3 \sqrt{2}} & -i\frac{t_2}{\sqrt{2}}  & \frac{t_3-t_1}{\sqrt{6}} & i\frac{t_2}{\sqrt{6}} \\
 0 & \frac{2 t_1+t_3}{3} & -i\frac{t_2}{\sqrt{2}} & \frac{t_3-t_1}{3 \sqrt{2}} & i\frac{t_2}{\sqrt{6}} & \frac{t_3-t_1}{\sqrt{6}} \\
\hline
\frac{t_3-t_1}{3 \sqrt{2}} & i\frac{t_2}{\sqrt{2}} & \frac{5t_1+t_3}{6} & 0 & \frac{t_3-t_1}{2 \sqrt{3}} & -i\frac{t_2}{\sqrt{3}} \\
 i\frac{t_2}{\sqrt{2}} & \frac{t_3-t_1}{3 \sqrt{2}} & 0 & \frac{5t_1+t_3}{6} & -i\frac{t_2}{\sqrt{3}} & \frac{t_3-t_1}{2 \sqrt{3}} \\
\hline
\frac{t_3-t_1}{\sqrt{6}} & -i\frac{t_2}{\sqrt{6}} & \frac{t_3-t_1}{2 \sqrt{3}} & i\frac{t_2}{\sqrt{3}} & \frac{t_1+t_3}{2} & 0 \\
-i\frac{t_2}{\sqrt{6}} & \frac{t_3-t_1}{\sqrt{6}} & i\frac{t_2}{\sqrt{3}} & \frac{t_3-t_1}{2 \sqrt{3}} & 0 & \frac{t_1+t_3}{2} \\ \end{array} \right), \nonumber \\
\ \\
\tilde {T}_{ij} ^{Y}
&=& \left( 
\begin{array}{cc|cc|cc}
\frac{2 t_1+t_3}{3} & 0 & \frac{t_3-t_1}{3 \sqrt{2}} & \frac{t_2}{\sqrt{2}} & \frac{t_1-t_3}{\sqrt{6}} & \frac{t_2}{\sqrt{6}} \\
0 & \frac{2 t_1+t_3}{3} & -\frac{t_2}{\sqrt{2}} & \frac{t_3-t_1}{3 \sqrt{2}} & -\frac{t_2}{\sqrt{6}} & \frac{t_1-t_3}{\sqrt{6}} \\
\hline
\frac{t_3-t_1}{3 \sqrt{2}} & -\frac{t_2}{\sqrt{2}} & \frac{5 t_1+t_3}{6} & 0 & \frac{t_1-t_3}{2 \sqrt{3}} & -\frac{t_2}{\sqrt{3}} \\
\frac{t_2}{\sqrt{2}} & \frac{t_3-t_1}{3 \sqrt{2}} & 0 & \frac{5 t_1+t_3}{6} & \frac{t_2}{\sqrt{3}} & \frac{t_1-t_3}{2 \sqrt{3}} \\
\hline
\frac{t_1-t_3}{\sqrt{6}} & -\frac{t_2}{\sqrt{6}} & \frac{t_1-t_3}{2 \sqrt{3}} & \frac{t_2}{\sqrt{3}} & \frac{t_1+t_3}{2} & 0 \\
\frac{t_2}{\sqrt{6}} & \frac{t_1-t_3}{\sqrt{6}} & -\frac{t_2}{\sqrt{3}} & \frac{t_1-t_3}{2 \sqrt{3}} & 0 & \frac{t_1+t_3}{2} \\ \end{array} \right)
\label{hopmat2}
\end{eqnarray}

The orbital mixing hopping term $t_2$ is absent in the pseudo-orbital diagonal blocks due to cancellation between the two O(Cl)-assisted hopping pathways. However, the $t_2$ term survives in the off-diagonal blocks, resulting in spin-dependent hopping term $i\sigma_\mu t_\mu$ for the $\mu$ =$X,Y,Z$ type bond, and consequent spin-rotation-symmetry breaking. 

\section{Magnetic interactions induced between NNN and NN sites}
For the planar zigzag order corresponding to parameter set B, the interaction terms between the $J=1/2$ sector, as evaluated from Eq. (\ref{couplings}) for the NNN sites $i,j$ and the three spin components $\alpha,\beta=x,y,z$, are given below for the three types of bonds: 

\begin{eqnarray}
& & [{\cal J}_{ij}^{\alpha\beta}]_Z ^{\frac{1}{2},\frac{1}{2}} = \left( \begin{array}{c c c} 
-0.046 & -0.011 & 0.001 \\
-0.023 & -0.046 & -0.020 \\
-0.020 & 0.002 & -0.091 
\end{array} \right ), \;\;\;
[{\cal J}_{ij}^{\alpha\beta}]_X ^{\frac{1}{2},\frac{1}{2}} = \left( \begin{array}{c c c} 
0.010 & -0.024 & -0.001 \\
-0.004 & 0.003 & -0.019 \\
0.023 & 0.013 & 0.008 
\end{array} \right ), \nonumber \\ \nonumber \\
& & [{\cal J}_{ij}^{\alpha\beta}]_Y ^{\frac{1}{2},\frac{1}{2}} = \left( \begin{array}{c c c} 
0.003 & -0.004 & -0.019 \\
-0.024 & 0.010 & -0.001 \\
0.013 & 0.022 & 0.008 
\end{array} \right ).
\label{intmat}
\end{eqnarray}
From the above interaction matrix elements, the exchange constants can be extracted by comparing with the standard forms of the Heisenberg ($J{\bf S}_i . {\bf S}_j$), Kitaev ($K^\gamma [S_i^\gamma S_j^\gamma - S_i^\alpha S_j^\alpha - S_i^\beta S_j^\beta$]), SOD ($\Gamma^{\alpha \beta}[S_i ^\alpha S_j ^\beta + S_i ^\beta S_j ^\alpha])$ and DM (${\bf D}.{\bf S}_i \times {\bf S}_j$) interactions. 

Similarly, the interaction terms for NN sites between the $J$=1/2 moment and the $J$=3/2 moment (in the pseudo orbital $l$=3) are obtained as below:
\begin{eqnarray}
& & [{\cal J}_{ij}^{\alpha \beta}]_Z ^{\frac{1}{2},\frac{3}{2}}
 = \left(
\begin{array}{c c c}
0.011 & 0.052 & 0.009 \\
0.052 & 0.011 & 0.009 \\
0.009 & 0.009 & -0.064
 \end{array} \right), \;\;\;
[{\cal J}_{ij}^{\alpha \beta}]_X ^{\frac{1}{2},\frac{3}{2}}
= \left(
 \begin{array}{c c c}
-0.029 & -0.099 & 0.089 \\
-0.128 & -0.023 & 0.022 \\
-0.016 & -0.096 & -0.130
 \end{array} \right), \nonumber \\ \nonumber \\
& &
[{\cal J}_{ij}^{\alpha \beta}]_Y ^{\frac{1}{2},\frac{3}{2}}
 = \left(
\begin{array}{c c c}
-0.023 & -0.128 & 0.022 \\
-0.099 & -0.029 & 0.089 \\
-0.096 & -0.016 & -0.130
 \end{array} \right).
\end{eqnarray}
The dominant NN interaction terms above are seen to frustrate the planar zigzag order for all three bonds, as also the interaction terms involving the other pseudo orbital ($l=2$).

\end{document}